\documentstyle[ltwol,epsfig]{article}

\def\PLB{{\em Phys. Lett.}  B}

\def\be{\begin{equation}}
\def\ee{\end{equation}}
\def\bea{\begin{eqnarray}}
\def\eea{\end{eqnarray}}

\begin{document}

\title{B/${\mathrm \bf \bar B}$ FLAVOUR TAGGING AND DOUBLY CHARMED B DECAYS IN ALEPH} 

\author{R. BARATE}

\address{LAPP, Chemin de Bellevue, BP110, F74941, Annecy-le-Vieux, 
CEDEX, France\\E-mail: barate@lapp.in2p3.fr}   


\twocolumn[\maketitle\abstracts{ This contribution concerns
three contributed papers that share the common feature
of analysing fully- (or almost fully-) reconstructed B decays
coming from a sample of four
million hadronic Z decays collected with the ALEPH detector at LEP.
In the first paper
\cite{pap1},
404 charged and neutral B mesons
decaying in standard modes
are fully reconstructed and used to look for
resonant structure in the B$\pi^{\pm}$ system.
In the framework of Heavy Quark Symmetry (HQS),
the mass of the ${\mathrm B_2^*}$ state and 
the relative production rate of the ${\mathrm B^{**}}$ system are measured.
In the same sample of B mesons, significant B$\pi^{\pm}$ charge-flavour
correlations are observed.
In the second paper
\cite{pap2},
a search for doubly-charmed B decays with both charmed mesons 
reconstructed is performed.
A clear signal is observed in the
 channels ${\mathrm b\rightarrow D_s \bar D}(X)$ and 
${\mathrm b\rightarrow D \bar D}(X)$   
 providing the first direct evidence for doubly-charmed b decays involving 
no ${\mathrm D_s}$ production. Evidence for associated ${\mathrm K^0_S}$ and 
${\mathrm K^{\pm}}$ production in the decays 
${\mathrm B\rightarrow D \bar D}(X)$ 
is also presented and some candidates for completely reconstructed B meson decays 
${\mathrm B\rightarrow D_s \bar D}(n\pi)$,
${\mathrm B\rightarrow D \bar D K^0_S}$ and 
${\mathrm B\rightarrow D \bar D K^\pm}$ are observed. 
 Furthermore, candidates for the two-body Cabibbo suppressed decays 
${\mathrm B^0\rightarrow D^{*-}D^{*+}}$ 
and 
${\mathrm B^-\rightarrow D^{(*)0}D^{(*)-}}$ are also observed.
One ${\mathrm B^0_s \rightarrow D_s^{+}D_s^{-}}$ 
event is reconstructed, which can be only the short-lived CP even
eigenstate.
In the third paper
\cite{pap3},
the ${\mathrm B_s}$ decay to ${\mathrm D_s^{(*)+}D_s^{(*)-}(X)}$ 
is observed, tagging the final state 
with two $\phi$ in the same hemisphere.
It corresponds mostly to the short-lived CP even eigenstate.
A preliminary value of the ${\mathrm B_s}$ short lifetime is obtained.}]

\section{Resonant Structure and Flavour Tagging in the B$\pi^{\pm}$ System}
\subsection{Introduction}\label{subsec:bint}
Fully reconstructed B meson decays are used 
to extract a precise mass of the
${\mathrm B_2^*}$ state and to obtain the B/${\mathrm \bar B}$ 
signature at the decay point.
Using the $\pi$ from
${\mathrm B^{**}}$ decay or the nearest $\pi$ from fragmentation,
direct tagging of the initial
B just before it oscillates and decays is possible. 
It could be used in future
CP violation experiments.

\subsection{${\mathrm B}$ meson and associated pion selection}\label{subsec:bsel}
Charged and neutral B mesons are fully reconstructed 
in various exclusive modes. 
Eighty percent are in the 
mode\footnote{Throughout this paper, charge conjugate decay modes are
always implied.}
${\mathrm B \rightarrow \bar D^*}(X)$, where $X$ is a
charged $\pi$, $\rho$ or ${\mathrm a_1}$, and 20\% are of the form 
B$\to J/\psi(\psi^\prime) X$, where $X$ is a charged $K$ or a neutral $K^*$. 
In addition, charged B candidates are reconstructed in the channels
${\mathrm B^- \rightarrow D^{*0}\pi^-}$
and ${\mathrm D^{*0} a_1^-}$, with a missing soft $\gamma$
or $\pi^0$ from the ${\mathrm D^{*0} \rightarrow D^{0}\gamma}$
or ${\mathrm D^0 \pi^0}$ decays.  
In total, 238 charged and 166 neutral B candidates are reconstructed with
purities of ($84\pm3\,$)$\%$ and ($86\pm3\,$)$\%$ respectively.

The neighbouring pion is selected using the $P_L^{max}$ algorithm
which chooses the track with the highest projected momentum
along the B direction (and a B$\pi$ mass below 7.3 GeV/$c$).

\subsection{Resonant Structure in the ${\mathrm B \pi^{\pm}}$ 
System}\label{subsec:bbw}
\begin{figure}[h]
\center
\mbox{\epsfig{file=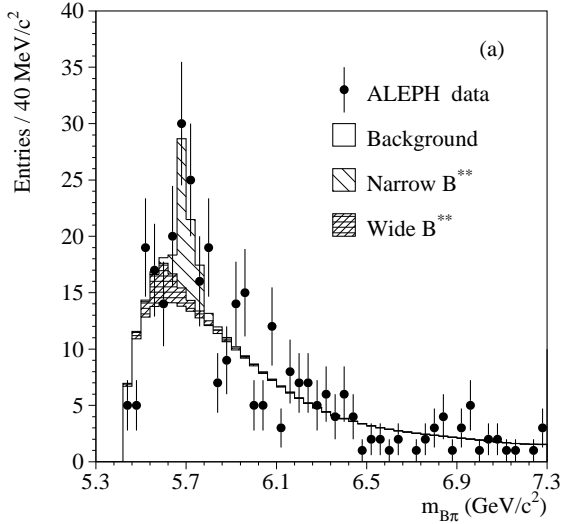,height=8cm}} \vspace*{-.5 cm}
\caption{The B$\pi$ mass spectrum from data (points with error bars)
and fit (histogram). The fit includes the expected
background plus contributions
from the narrow and wide ${\mathrm B^{**}}$ states.}
\label{ffitb}
\end{figure}

Using the pion selected with the $P_L^{max}$ algorithm,
the right sign and wrong sign B$\pi$ mass
distributions are made
and ${\mathrm B^{**}}$ signals are extracted.
The gain of one order of magnitude in mass resolution
compared to previous inclusive experiments,
due to the quality
of exclusive B decays, allows a more precise measurement of the masses
if one uses the Heavy Quark Symmetry parameters
\cite{pap1}.
HQS predicts
4 resonances giving
5 correlated Breit Wigner (3 narrow and 2 wide) in the
B$\pi$ mass distributions.
Here only the overall mass scale and the total number of signal events are
left free.
An unbinned likelihood fit 
(Fig.~\ref{ffitb})
gives :
$$M(B_2^*) = (5739^{+\ 8}_{-11}\mathrm{(stat)}^{+6}_{-4}\mathrm{(syst)}) 
\,\mathrm{MeV/c}^2$$
$$f_{B^{**}}\equiv
\frac{{\cal B}(b \rightarrow B^{**} \rightarrow B^{(*)}\pi)}{{\cal B}(b \rightarrow
B_{u,d})} = (31\pm 9\mathrm{(stat)}^{+6}_{-5}\mathrm{(syst)})\%.$$
The result for the mass of the ${\mathrm B_2^*}$ state is somewhat low compared to the
predicted value of 5771\,MeV/$c^2$
.

\subsection{${\mathrm B/ \bar B}$ Flavour Tagging}\label{subsec:bft}
The sign of the neighbouring pion (from ${\mathrm B^{**}}$ or fragmentation)
tags the B/${\mathrm \bar B}$ nature at production : 
a $\pi^+$ right-sign tags a ${\mathrm B^0}$, etc.
The $P_L^{max}$ tagging algorithm efficiencies 
for neutral and charged B's are :
\[ \epsilon_{tag}^{N}=(89\pm 3(\mathrm {stat})\pm 2(\mathrm {syst}))\,\%\,, \]
\[ \epsilon^C_{tag}=(89\pm 2(\mathrm {stat})\pm 1(\mathrm {syst}))\,\%. \]

\begin{figure}[h]
\center
\mbox{\epsfig{file=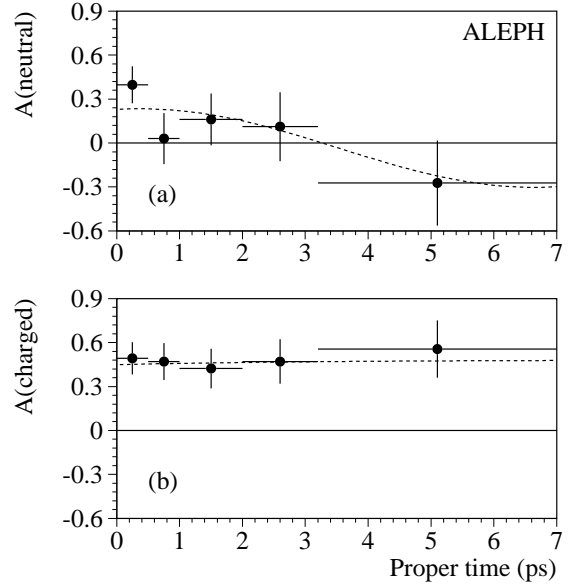,height=9cm}} \vspace*{-1.0 cm}
\caption{The right-sign/wrong-sign asymmetries in the data as a function of the
proper decay time. The dashed curves display the charge asymmetries
determined from the unbinned likelihood fits.}
\label{fasym}
\end{figure}

The asymmetry ${\mathcal A}^N$ between the number of right-sign and 
wrong-sign tags, 
${\mathcal A}^N = (N_{rs}-N_{ws})/(N_{rs}+N_{ws})$, 
is shown as a function of the B decay proper time, $t$, in 
Fig.~\ref{fasym}a. 
The sinusoidal mixing of ${\mathrm B^0}$
into ${\mathrm \bar B^0}$ gives rise to the excess of 
wrong-sign tags (negative value of ${\mathcal A}^N$) at high proper times.

The mistag
rate for neutral B's, $\omega^N_{tag}$,
is measured from the oscillation amplitude, with 
$\Delta m_d$ fixed to the world 
average of  $0.474\pm 0.031\,$ps$^{-1}$
.
 An unbinned likelihood fit gives:
\[ \omega^N_{tag}=(34.4 \pm 5.5(\mathrm {stat})\pm 1.0(\mathrm {syst}))
\,\%\,; \]
similarly, an unbinned likelihood fit for
$\omega^C_{tag}$, the mistag
rate for charged B's gives : 
\[ \omega^C_{tag}=(26.0\pm 3.6 (\mathrm {stat})\pm 0.7(\mathrm {syst}))\,\%\]
showing that this method
gives good tagging performance,
whilst being very efficient.
The corresponding fits to the data are displayed in Fig. 2.

\section{Doubly charmed B decays}
\subsection{Introduction}\label{subsec:dint}
The final states with 2 charm mesons allow a precise study of the $b$ quark
decay into $c \bar c s$ and give a direct access to the 
average number of charm quarks per $b$ decay
$n_c$.
The study of 3-body B meson decay in ${\mathrm \bar D D K}$ 
gives many results on the diagrams,
possible resonances and dynamics. 
Added to
the Cabbibo suppressed ${\mathrm D \bar D}$,
standard ${\mathrm D_s^+\bar D}$, 
and new ${\mathrm D_s^+\bar D }n\pi^\pm$ decays, 
one obtains the total double
charm contribution to B meson decay.

\subsection{${\mathrm D\bar D}$ selection}\label{subsec:dsel}
The charmed mesons are searched for in the decay modes 
${\mathrm  D^0\rightarrow K^-\pi^+}$, 
${\mathrm  D^0\rightarrow K^-\pi^+\pi^-\pi^+}$,
${\mathrm  D^+\rightarrow K^-\pi^+\pi^+}$,     
${\mathrm  D^{*+}\rightarrow D^0 \pi^+}$,     
${\mathrm  D_s^+\rightarrow \phi\pi^+ (\phi\rightarrow K^-K^+)}$ 
and ${\mathrm  D_s^+\rightarrow \bar K^{*0}K^+}$ 
 ${\mathrm(\bar K^{*0}\rightarrow K^-\pi^+)}$. For 
${\mathrm  D^0}$ mesons from ${\mathrm  D^{*+}}$ decay, the decay mode 
${\mathrm  D^0\rightarrow K^-\pi^+\pi^0}$ is also used.

The pairs of D candidates must be in
the same hemisphere
and the two D candidates are required to form a vertex 
with a probability of at least 0.1\%.
To maintain a good acceptance for the 
${\mathrm B\rightarrow D\bar D}X$ signal whilst rejecting the backgrounds and 
minimizing the model dependence of the selection efficiencies, a cut 
$d_{{\mathrm BD}}/\sigma_{{\mathrm BD}}>-2$ ($>$0) is applied on the D$^0$, 
${\mathrm D_s^+}$ (D$^+$) decay length significance (defined in Fig. 3).
The decay length significance of the 
${\mathrm D\bar D}$ vertex is also required to satisfy the condition 
$d_{{\mathrm B}}/\sigma_{{\mathrm B}}>-2$.

To obtain the number of real ${\mathrm D\bar D}$ events,
standard tables are made (for instance $m(K\pi)_1 / m(K\pi)_2$) 
and the combinatorial background is subtracted linearly.

\subsection{Inclusive b quark decays in ${\mathrm D_s \bar D} (X)$ or
${\mathrm D \bar D} (X)$}\label{subsec:dbq}
After acceptance corrections, the different branching fractions 
obtained are given in
Table 1.

The inclusive branching fraction of b quarks to 
${\mathrm D_s D} (X)$ is measured to be  
 $$ {\mathrm {\cal B}(b \rightarrow D_s D^0, D_s D^\pm} (X))\!=\!{\mathrm 
\left (13.1^{+2.6}_{-2.2} (st) ^{+1.8}_{-1.6} (sy) 
       ^{+4.4}_{-2.7}({\cal B}_D) \right )}\!\%,$$
 in good agreement with previous measurements of the inclusive branching 
 fraction of the B mesons to ${\mathrm D_s}$
 .

\begin{figure}[h]
\center
\mbox{\epsfig{file=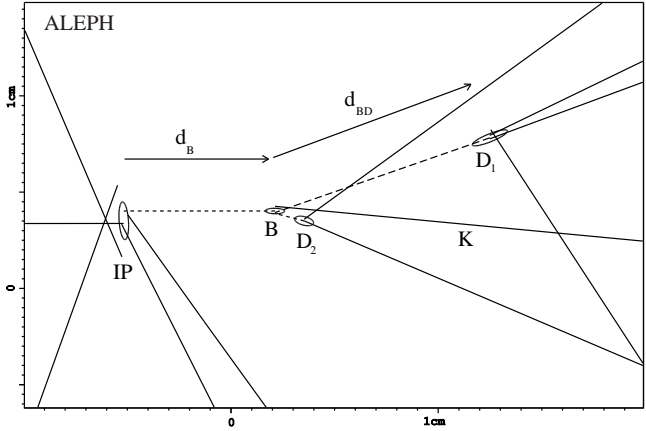,height=5.7cm}} \vspace*{-0.5 cm}
\caption{Display of a decay ${\mathrm B^0\rightarrow D^- D^0 K^+}$ 
reconstructed in the ALEPH detector (real data).}
\label{fevent}
\end{figure}

\begin{table}[h]
\begin{center}
\caption{Summary of the different branching fractions 
measured in this analysis.
  The modes involving a ${\mathrm D^{*+}}$ (lowest part of the table) 
are also included in the upper part results as a subsample of the modes 
involving a ${\mathrm D^0}$ or a ${\mathrm D^+}$.}
\label{tb}
\vspace{0.2cm}
\begin{tabular}{|c|c|}
\hline  
   Channel  & ${\cal B}$(\%)  \\ \hline 
 ${\mathrm  b\rightarrow D^0 D_s^-} (X) $ 
& $9.1^{+2.0}_{-1.8}\ ^{+1.3}_{-1.2} \ ^{+3.1}_{-1.9}$  \\
 ${\mathrm  b\rightarrow D^+ D_s^-} (X) $ 
& $4.0^{+1.7}_{-1.4}\pm 0.7  \ ^{+1.4}_{-0.9}$  \\ \hline 
 Sum 
 ${\mathrm  b\rightarrow D^0 D_s^- , D^+ D_s^- }(X)$ 
&$13.1^{+2.6}_{-2.2}\ ^{+1.8}_{-1.6}\ ^{+4.4}_{-2.7}$\\ \hline 
 ${\mathrm  b\rightarrow D^0 \bar D^0 }(X)$  
& $5.1^{+1.6}_{-1.4}\ ^{+1.2}_{-1.1} \pm 0.3 $ \\
 ${\mathrm  b\rightarrow D^0 D^-,D^+ \bar D^0}(X) $ 
& $2.7^{+1.5}_{-1.3}\ ^{+1.0}_{-0.9}  \pm 0.2 $ \\ 
 ${\mathrm  b\rightarrow D^+ D^-}(X)$   
& $<0.9\%$ at 90\%C.L. \\ \hline 
 Sum ${\mathrm  b\rightarrow D^0 \bar D^0,D^0 D^-,D^+ \bar D^0}(X)$ 
& $7.8_{-1.8}^{+2.0}\ _{-1.5}^{+1.7} \ _{-0.4}^{+0.5}$\\ \hline 
 ${\mathrm  b\rightarrow D^{*+} D_s^{-} }(X)$ 
& $3.3^{+1.0}_{-0.9} \pm 0.6 \ ^{+1.1}_{-0.7} $ \\
 ${\mathrm  b\rightarrow D^{*+} \bar D^0 , D^0D^{*-}}(X) $
& $3.0^{+0.9}_{-0.8}\ ^{+0.7}_{-0.5} \pm 0.2 $ \\
 ${\mathrm  b\rightarrow D^{*+}D^-,D^+D^{*-}}(X)$ 
& $2.5^{+1.0}_{-0.9}\ ^{+0.6}_{-0.5} \pm 0.2 $ \\ \hline 
 ${\mathrm  b\rightarrow D^{*+}      D^{*-}}(X)$
& $1.2 ^{+0.4}_{-0.3}\pm 0.2 \pm 0.1 $ \\ \hline 
\end {tabular}
\end{center}
\end{table}

 For the first time, doubly-charmed B decays involving no ${\mathrm D_s}$ 
 production are observed. The corresponding inclusive branching 
 fractions are  
  $$ {\mathrm {\cal B}(b \rightarrow D^0 \bar D^0, D^0 D^\pm} (X) )
 \!=\!{\mathrm 
 \left ( 7.8^{+2.0}_{-1.8}(st) ^{+1.7}_{-1.5}(sy) 
         ^{+0.5}_{-0.4} ({\cal B}_D)\right )\!\% }$$
Hence a significant fraction of the 
doubly-charmed B decays leads to no ${\mathrm D_s}$ production. For the average 
mixture of b hadrons produced at LEP, the sum over all the decay modes above 
yields:
 $$ {\mathrm {\cal B}
(b \rightarrow D_s D^0, D_s D^\pm, D^0 \bar D^0, D^0 D^\pm} (X)) = 
$$     
 $$ {\mathrm \left (20.9^{+3.2}_{-2.8} (stat) ^{+2.5}_{-2.2} (syst) 
       ^{+4.5}_{-2.8}({\cal B}_D) \right )}\%.$$
Adding the small hidden charm and charmed baryon contributions,
this measurement is in good agreement 
with the recent ALEPH measurement of the
average number of charm quarks per $b$ decay
:

$n_c={\mathrm 1.230 \pm 0.036(stat) \pm 0.038(syst) \pm 0.053({\cal B}_D)}$
.
Some corresponding ${\mathrm D^{(*)} \bar D^{(*)}}$ mass spectra
are given in Fig. 4.

\begin{figure}[h]
\center
\mbox{\epsfig{file=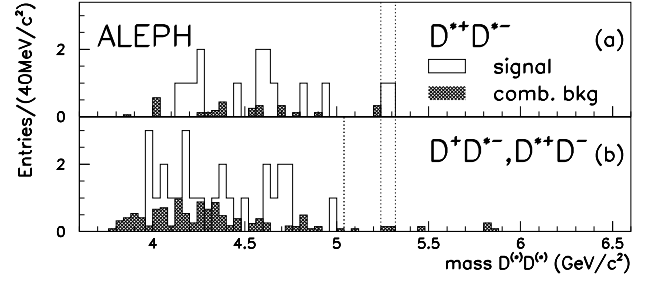,height=11cm}} \vspace*{-7 cm}
\caption{Unshaded histogram: the ${\mathrm D\bar D}$ mass spectra of the selected 
${\mathrm b \rightarrow D\bar D}(X)$ candidates (a) ${\mathrm D^{*+} D^{*-}}$ 
(b) ${\mathrm D^{\pm } D^{*\mp}}$ . The channels are mutually exclusive. Shaded histogram:  
  the ${\mathrm D\bar D}$ mass distribution of the events in the sidebands
 of the  ${\mathrm D_1}$  or  ${\mathrm D_2}$ mass spectra, normalised to 
the expected number of 
combinatorial background events.}
\label{fdd}
\end{figure}

\subsection{Cabbibo suppressed ${\mathrm B\rightarrow \bar D D}$ decays}\label{subsec:dcab}
As can be seen in Fig. 4a,
two candidates for the Cabibbo suppressed  
decay ${\mathrm B^0_d \rightarrow D^{*+} D^{*-}}$ 
are observed. 
Asking for no charged track at the ${\mathrm D\bar D}$ vertex,
the background is reduced to $0.10\pm 0.03$ event.
The corresponding branching fraction is measured to be
$${\mathrm {\cal B}(\bar B^0_d \rightarrow D^{*+} D^{*-}) = \left (
 0.23_{-0.12}^{+0.19}\pm 0.04\pm 0.02({\cal B}_D) \right )\% .}$$
One candidate for the Cabibbo suppressed decay 
${\mathrm B^- \rightarrow D^{*-} D^0}$, with both D vertices well separated from
the reconstructed B decay point, is also observed
\cite{disp}
and limits on branching fractions are obtained
\cite{pap2}.

\subsection{Semi inclusive ${\mathrm B\rightarrow \bar D D K}(X)$ decays}\label{subsec:dsi}
Reconstructing a
${\mathrm K^0_S}$ and ${\mathrm K^\pm}$
compatible with the
${\mathrm D \bar D}$ vertex,
and making the ${\mathrm \bar D D K}$ invariant mass 
(without using the soft pion from a ${\mathrm D^*}$),
3 peaks separated by about 150 MeV/$c^2$ appear,
corresponding to the three 3-body B meson decay modes (Fig. 5).
At a lower mass, the events correspond to more than 3-body decay modes 
(mainly ${\mathrm D^{(*)} \bar D^{(*)} K \pi}$).
After background subtraction, the total number of events is
32.2$\pm$7.9,
and in the 3-body region
21.2$\pm$5.5.
Hence, one sees that the three-body 
decays ${\mathrm B \rightarrow \bar D^{(*)} D^{(*)} K}$ are a large part
(about $70\%$)
of the inclusive doubly-charmed 
${\mathrm B \rightarrow \bar D^{(*)} D^{(*)} K (X)}$ decays.

\subsection{Exclusive ${\mathrm B\rightarrow \bar D D K}$ decays}\label{subsec:d3}
Asking now for no other charged track at the ${\mathrm \bar D D K}$ vertex 
(and using the soft pion from a ${\mathrm D^*}$ when available),
a clear B signal appear (Fig. 6) with 
9 ${\mathrm \bar D^{(*)} D^{(*)} K^0_S}$ events
and 9 ${\mathrm \bar D^{(*)} D^{(*)} K^\pm}$ events above 5.04 GeV/$c^2$.
One of these events
\cite{disp}
was displayed in Fig. 3.

No evidence for resonant 
decays ${\mathrm B \rightarrow \bar D^{(*)} D^+_{s1}(2535) }$  is found.

\begin{figure}[h]
\center
\mbox{\epsfig{file=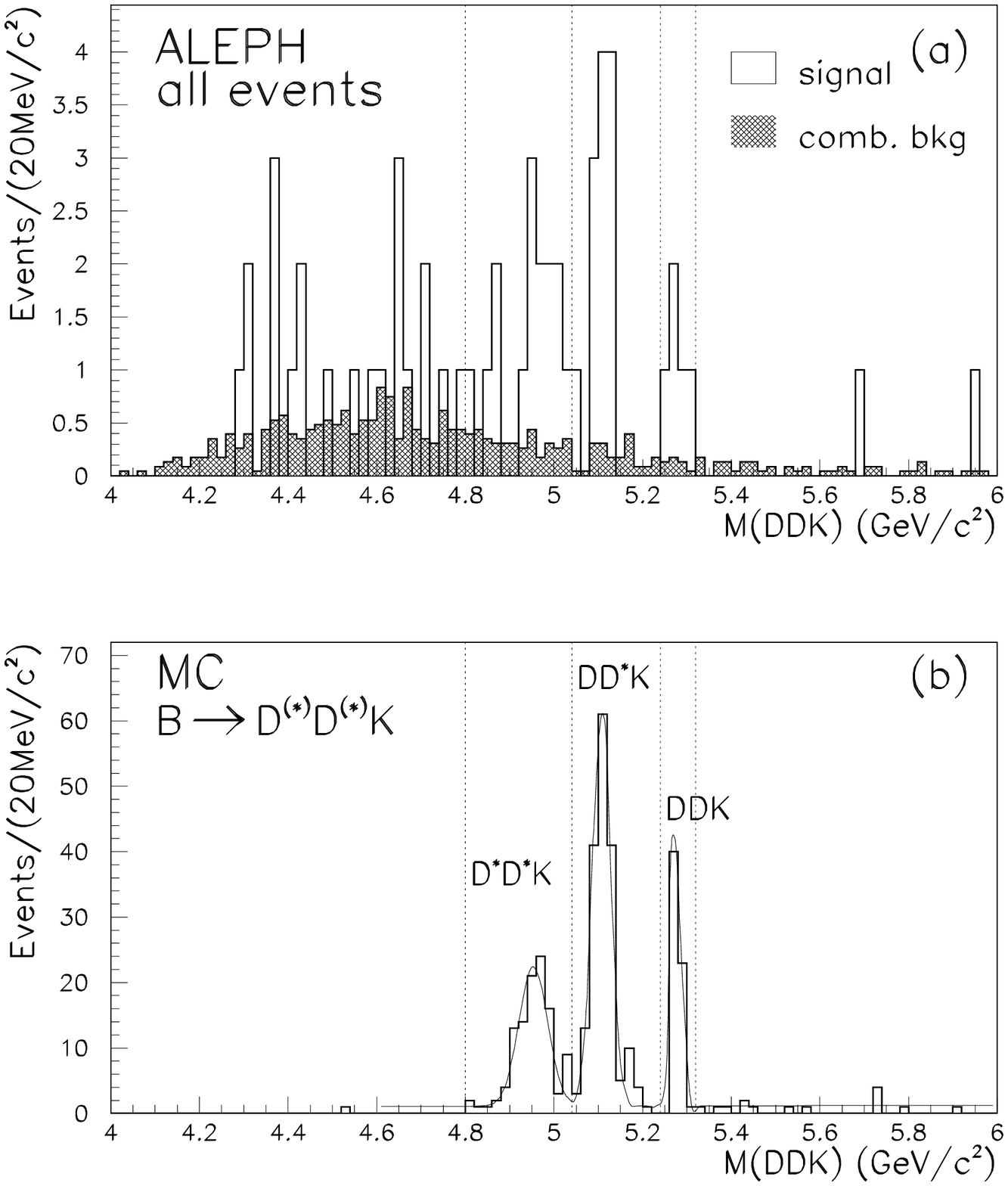,height=10cm}} \vspace*{-0.5 cm}
\caption{The ${\mathrm  D^0 \bar D^0 K}$,
${\mathrm D^0D^-K}$ or 
${\mathrm D^+D^-K}$ mass of ${\mathrm  D \bar D}$ events with a reconstructed 
${\mathrm K^0_S}$ or a K$^\pm$ for  (a) ALEPH data
(b) simulated three-body decays 
${\mathrm B \rightarrow D^{(*)} \bar D^{(*)} K}$. The $\pi^+$ from 
${\mathrm D^{*+}\rightarrow D^0 \pi^+}$, even if reconstructed, are not used in 
the mass.}
\label{fddkx}
\end{figure}

${\mathrm B}$ and ${\mathrm \bar B}$ decays give different final states
and, for instance in a ${\mathrm \bar B}$ decay,
the D coming from the b quark decay (called ${\mathrm D_b}$) can be distinguished
from the ${\mathrm \bar D}$ coming from the W decay (${\mathrm D_W}$):
in most of the events,
the invariant mass $m{\mathrm (D_bK)}$ tend to be 
higher than  $m{\mathrm (D_WK)}$ (and hence the momentum 
$p({\mathrm D_b})$ in the B rest frame is higher than  $p({\mathrm D_W})$).

The diagrams contributing to these decays can be divided
into 3 classes (Table 2):
External W (E),
Internal W (I) or color suppressed
(cf ${\mathrm B \rightarrow J/\psi K}$)
and interference (EI); as can be seen in Table 2,
some I type events are also observed. 

Using isospin symmetry, the different branching fractions are given in Table 2
and the sum is :
$${\cal B}{\mathrm (B \rightarrow \bar D^{(*)} D^{(*)} K) = \left (
 7.1_{-1.5}^{+2.5}(st) _{-0.8}^{+0.9}(sy) \pm 0.5({\cal B}_D) 
\right )\!\%. }$$

\subsection{${\mathrm B\rightarrow \bar D D K X}$ decays}\label{subsec:d4}
Compared
to the semi-inclusive result of
Sec.~\ref{subsec:dsi}
or to the inclusive b results of
Sec.~\ref{subsec:dbq}
, scaled by a factor 
${\mathrm 1/2f_{B^0_d}}=1.3$ to 
account for ${\mathrm b \rightarrow \bar B^0,B^-}$, 
one sees that ${\cal B}{\mathrm (B \rightarrow \bar D^{(*)} D^{(*)} K X)}$  
should be about $3\%$.

\begin{figure}[h]
\center
\mbox{\epsfig{file=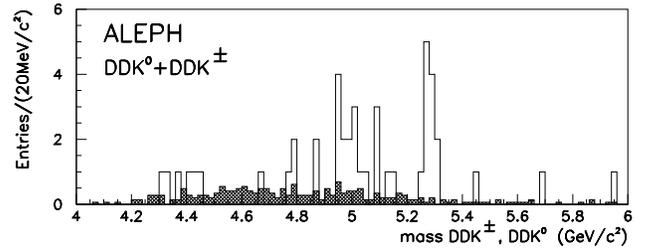,height=11cm}} \vspace*{-7.5 cm}
\caption{Invariant mass $m({\mathrm D\bar DK})$ 
for events with one identified K and no other additional track from 
the ${\mathrm D\bar DK}$ vertex. 
D can be either a D$^0$, a D$^+$ or a D$^{*+}$.}
\label{fddk}
\end{figure}

\begin{figure}[h]
\center
\mbox{\epsfig{file=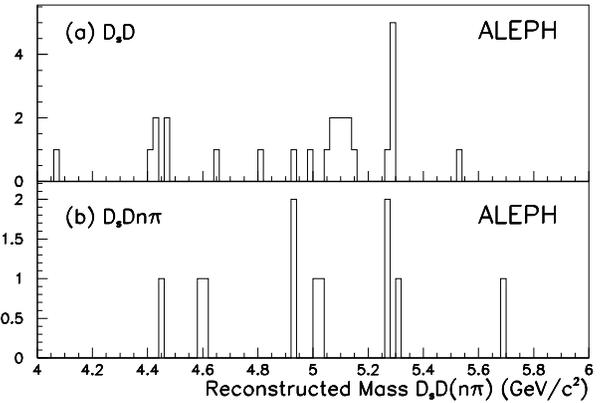,height=11cm}} \vspace*{-5.5 cm}
\caption{Invariant mass $m({\mathrm D_s^+ \bar D}(n\pi^\pm))$ 
 reconstructed for 
ALEPH data  
(a) ${\mathrm D_s^+\bar D}$
(b) ${\mathrm D_s^+\bar D}n\pi^\pm$, $n\ge 1$.
The peak close to 5.1 GeV/$c^2$ is due to events with one missing neutral from 
decays ${\mathrm D^{*}\rightarrow D \pi^0,\gamma}$ or 
or ${\mathrm D_s^{*+}\rightarrow D_s^{+}\gamma}$.}
\label{fdsd}
\end{figure}

\begin{table*}[t]
\caption{Summary of the various branching fractions 
 ${\mathrm B\rightarrow D \bar D K}$ measured in this analysis. 
\label{tddk}}
\vspace{0.2cm}
\begin{center}
\begin{tabular}{|c|c|c|c|}
\hline  
 Diagram & Channel        & Number of & 
 ${\cal B}({\mathrm B\rightarrow D^{(*)} \bar D^{(*)} K})$ \\ 
         & (B$^0$, B$^+$) & candidates& 
 (B$^0$/B$^+$ average)    \\
\hline 
E & ${\mathrm D^-D^0K^+}$, ${\mathrm \bar D^0 D^+K^0}$  & 3 & 
    $1.7^{+1.2}_{-0.8}\pm 0.2\pm 0.1$\% \cr
E & ${\mathrm (D^{*-}D^0+D^-D^{*0})K^+}$, 
    ${\mathrm (\bar D^{*0} D^++\bar D^{0} D^{*+})K^0}$ & 5 &
    $1.8_{-0.8}^{+1.0}\pm 0.3\pm 0.1$\% \cr
E & ${\mathrm D^{*-}D^{*0}K^+}$, ${\mathrm \bar D^{*0} D^{*+}K^0}$& 1 &
    $<1.3\%$ \cr 
\hline 
I & ${\mathrm \bar D^0D^0K^0}$, ${\mathrm D^+D^-K^+}$ & 1 & 
    $<2.0\%$ \cr
I & ${\mathrm (\bar D^0D^{*0}+\bar D^{*0}D^0)K^0}$,    
    ${\mathrm (D^{*+}D^-+D^{+}D^{*-})K^+}$ & 1 & 
    $<1.6\%$ \cr
I & ${\mathrm \bar D^{*0}D^{*0}K^0}$, ${\mathrm D^{*+}D^{*-}K^+}$ & 1 &
    $<1.5\%$ \cr
\hline 
EI & ${\mathrm D^+D^-K^0}$, ${\mathrm \bar D^0D^0K^+}$ & 1 &
    $<1.9\%$ \cr
EI & ${\mathrm (D^{*+}D^-+D^+D^{*-})K^0}$, 
     ${\mathrm (\bar D^{*0}D^0+\bar D^0D^{*0})K^+}$ & 4 &
     $1.6_{-0.7}^{+1.0}\pm 0.2 \pm 0.1$\% \cr 
EI & ${\mathrm D^{*+}D^{*-}K^0}$, ${\mathrm \bar D^{*0}D^{*0}K^+}$ & 1 &
     $<3.0$\% \cr 
\hline\hline
Sum E  & ${\mathrm D^{(*)-}D^{(*)0}K^+}$, ${\mathrm \bar D^{(*)0}D^{(*)+}K^0}$ & 9 &
     $3.5_{-1.1}^{+1.7}\ _{-0.4}^{+0.5}\pm 0.2$\% \cr 
Sum I  & ${\mathrm \bar D^{(*)0}D^{(*)0}K^0}$, ${\mathrm D^{(*)+}D^{(*)-}K^+}$ & 3 &
     $0.8_{-0.4}^{+1.0}\ _{-0.1}^{+0.2}\pm 0.1$\% \cr 
Sum EI & ${\mathrm D^{(*)+}D^{(*)-}K^0}$, ${\mathrm \bar D^{(*)0}D^{(*)0}K^+}$ & 6 &
     $2.8_{-1.0}^{+1.6}\ _{-0.3}^{+0.4}\pm 0.2 $\% \cr 
\hline\hline 
E+I+EI & Sum ${\mathrm D \bar DK}$ & 5 &
         $2.3^{+1.5}_{-0.9}\ ^{+0.3}_{-0.3}\pm 0.2$\%\cr
E+I+EI & Sum ${\mathrm D \bar D^*K+D^* \bar DK}$ & 10 &
         $3.8^{+1.6}_{-1.1}\ ^{+0.5}_{-0.4}\pm 0.2$\%\cr
E+I+EI & Sum ${\mathrm D^* \bar D^*K}$ & 3 &
         $1.0^{+1.3}_{-0.6}\ ^{+0.2}_{-0.2}\pm 0.1$\%\cr\hline
\hline
E+I+EI & Sum ${\mathrm D^{(*)} \bar D^{(*)}K}$ & 18 &
         $7.1^{+2.5}_{-1.5}\ ^{+0.9}_{-0.8}\pm 0.5$\%\cr \hline
\end {tabular}
\end{center}
\end{table*}

\subsection{${\mathrm B\rightarrow D_s^+\bar D }(n\pi^\pm)$ decays}
\label{subsec:dds}
Using the events with a ${\mathrm D_s^+\bar D}$ mass
above 5.04 GeV/$c^2$ in Fig. 7a,
the branching fraction of ${\mathrm B^0}$ and ${\mathrm B^+}$ 
mesons into doubly-charmed two-body decay modes is also measured and gives 
$$ {\cal B}({\mathrm B \rightarrow D_s^{(*)+} \bar D^{(*)}}) = {\mathrm
\left ( 5.6^{+2.1}_{-1.5}(st)\ ^{+0.9}_{-0.8}(sy) \ ^{+1.9}_{-1.1} 
({\cal B}_D)\right )}\%,$$
 in good agreement with previous measurements of the same quantity 
.

For the first time, some candidates for  
completely reconstructed  decays ${\mathrm B^0,B^+\rightarrow\bar D^{(*)}D_s^+}n\pi^\pm$ 
($n\ge 1$) are also observed (Fig. 7b). 
A measurement of the branching 
 fraction for many-body decays ${\mathrm B^0,B^+\rightarrow\bar D^{(*)}D_s^+}X$ 
is performed, leading to 
$$ {\cal B}({\mathrm B\rightarrow D_s^{(*)\pm} D^{(*)}}X) = {\mathrm
 \left ( 9.4^{+4.0}_{-3.1}(st) ^{+2.2}_{-1.8}(sy) 
        ^{+2.6}_{-1.6}({\cal B}_D)  \right )}\%.$$

\subsection{Conclusion on ${\mathrm B}$ meson decays}\label{subsec:dc}
Summing the results of
Sec.~\ref{subsec:dcab},
~\ref{subsec:d3},
~\ref{subsec:d4} and
~\ref{subsec:dds},
one sees that
${\mathrm B\rightarrow D\bar D }(X)$
and
${\mathrm B\rightarrow D_s^+\bar D }(X)$
are a big part (about $25\%$) of B mesons decays.

\subsection{${\mathrm B^0_s \rightarrow D_s^{+}D_s^{-}}$ 
decay}\label{subsec:dsds}
One event
\cite{disp}
is reconstructed with a ${\mathrm D_s^{+}D_s^{-}}$ mass 
at the ${\mathrm B^0_s}$ mass, on a negligible
background, with ${\mathrm  D_s^+\rightarrow \bar K^{*0}K^+}$ 
and ${\mathrm  D_s^-\rightarrow \phi\pi^-}$.
This can be only the pure CP even ${\mathrm B^0_s}$ short state
as explained in
Sec.~\ref{subsec:sint}
below. 

\section{Width difference in the ${\mathrm B_s - \bar{B_s}}$ system}
\subsection{Introduction}\label{subsec:sint}
Most of the channels common
to ${\mathrm B_s}$ and ${\mathrm \bar{B_s}}$ are CP even 
(cf ${\mathrm D_s^{+}D_s^{-}}$ ...).
Hence the ${\mathrm B_s}$ short is the CP even state.
Here a direct measurement of the ${\mathrm B_s}$ short lifetime is made 
using the mostly CP even decay modes
${\mathrm B^0_s \rightarrow D_s^{(*)+}D_s^{(*)-} (X)}$ 
with ${\mathrm  D_s^+\rightarrow \phi X}$
, that is a $\phi\phi X$ final state
\cite{rib}
.

\subsection{$\phi\phi$ selection}\label{subsec:ssel}
The same method as for double charm (Sec.~\ref{subsec:dsel}) is applied
(table $m(K^+ K^-)_1 / m(K^+ K^-)_2$).
The number of ${\mathrm \phi\phi}$ events measured in this way 
is $N_{\phi\phi} = 50 \pm 15$,
taking into account the non linear
background shape near the $K^+ K^-$ threshold.
In these events, $20\%$ of charm contamination is expected from Monte Carlo;
the contribution of $\phi\phi$ coming
from ${\mathrm B_d}$ and ${\mathrm B_u}$ decays
is evaluated
using $ {\cal B}({\mathrm B\rightarrow D_s^{(*)\pm} D^{(*)}}(X))$ 
measured in Sec.~\ref{subsec:dds}.
After subtracting these events, the number 
of ${\mathrm B^0_s \rightarrow D_s^{(*)+}D_s^{(*)-}(X)}$ found is
$N_{sig} = 32 \pm 17$
(over a background of  $78 \%$).
This number corresponds to a 
${\mathrm B^0_s \rightarrow D_s^{(*)+}D_s^{(*)-}(X)}$ branching 
ratio of approximately $10\%$.

\subsection{${\mathrm B_s}$ short lifetime}\label{subsec:stau}
The ${\mathrm \phi\phi}$ vertex 
(which has a resolution of about $200 \mu m$) 
is a good approximation of the ${\mathrm B_s}$ vertex 
due to the short ${\mathrm D_s}$ lifetime.
An unbinned maximum likelihood fit
using background parametrisation from the sidebands gives 
a preliminary lifetime 
$\tau_s = 1.42 \pm 0.23 \pm 0.16$ ps for this eigenstate.
Using the world average ${\mathrm B_s}$ lifetime, 
$\bar{\tau} = 1.61 \pm 0.10$~ps,
it would  correspond
\cite{rib}
to $\Delta\Gamma/\Gamma = (24 \pm 35)\%$ 
where the statistical and systematic errors are combined.

\end{document}